\documentclass[letter,traditabstract]{aa}  

\usepackage{graphicx}
\usepackage{txfonts}
 \usepackage{relsize}

\newcommand{\deriv} [2] {\frac {\textrm{d} #1 } {\textrm{d} #2} }
\newcommand{\derivt} [2] {\frac {\textrm{D} #1 } {\textrm{D} #2} }
\newcommand{\eq}[1]{Eq.~(\ref{#1})}
\newcommand{\eqs}[1]{Eqs.~(\ref{#1})}

\begin{document}

   \title{Surface effects and turbulent pressure}
   \subtitle{Assessing the Gas-$\Gamma_1$ and Reduced-$\Gamma_1$ empirical models}

   \author{K. Belkacem\inst{1} \and F. Kupka\inst{2,3,4} \and J. Philidet\inst{1} \and R. Samadi\inst{1}}

   \institute{LESIA, Observatoire de Paris, Université PSL, CNRS, Sorbonne Université, Université de Paris, 5 place Jules Janssen, 92195 Meudon, France, 
              \email{kevin.belkacem@obspm.fr}
              \and Univ. of Appl. Sciences Technikum Wien, Dept. Applied Mathematics and Physics, H\"ochst\"adtplatz 6, A-1200 Wien, Austria
              \and Wolfgang-Pauli-Institute c/o Faculty of Mathematics, University of Vienna, Oskar-Morgenstern-Platz 1, A-1090 Wien, Austria
              \and Max-Planck-Institut f\"{u}r Sonnensystemforschung, Justus-von-Liebig Weg 3, 37077 G\"{o}ttingen, Germany
             }

   \date{}
    
  \abstract
   {
   The use of the full potential of stellar seismology is made difficult by the improper modeling of the upper-most layers of solar-like stars and their influence on the modeled frequencies.  Our knowledge on these \emph{surface effects} has improved thanks to the use of 3D hydrodynamical simulations but the calculation of eigenfrequencies relies on empirical models for the description of the Lagrangian perturbation of turbulent pressure: the reduced-$\Gamma_1$ model (RGM) and the gas-$\Gamma_1$ model (GGM). Starting from the fully  compressible turbulence equations, we derive both the GGM and RGM models using a closure to model the flux of turbulent kinetic energy. It is found that both models originate from two terms: the source of turbulent pressure due to compression produced by the oscillations and the divergence of the flux of turbulent pressure. It is also  demonstrated that they are both compatible with the adiabatic approximation but also imply a number of questionable assumptions mainly regarding mode physics. Among others hypothesis, one has to neglect the Lagrangian perturbation of the dissipation of turbulent kinetic energy into heat and the Lagrangian perturbation of buoyancy work.  
   }


   \maketitle
%

\section{Introduction}

Systematic differences between observed and modeled eigenfrequencies is a long-standing problem in  stellar seismology. For the Sun and solar-like stars, this  has been identified and recognized to be associated to the improper modeling of their uppermost layers \cite[e.g.][]{Brown1984,Dziembowski1988,JCD1997}. To circumvent this problem, well-chosen combinations of frequencies \citep[e.g.][]{Roxburgh2003} or ad-hoc corrections \citep[][]{Kjeldsen2008,Ball2014,Sonoi2015} are commonly used. Nevertheless, to be able to exploit all the information contained in the observed frequencies, the physics underlying \emph{surface effects} must be understood. 

To do so, \cite{Rosenthal1999} used 3D hydrodynamical simulations. This allowed to account for the mean turbulent pressure \citep[as well as convective backwarming, see][]{Trampedach2017} in the equilibrium structure. Their work was followed by \cite{Piau2014,Magic2016,Houdek2017,Jorgensen2019,Schou2020} for the Sun and \cite{Sonoi2015,Ball2016,Trampedach2017,Sonoi2017,Jorgensen2017,Jorgensen2018,Manchon2018,Jorgensen2019b,Jorgensen2019c,Houdek2019,Mosumgaard2020,Jorgensen2021} for solar-like stars. A drawback of this approach is the need to compute adiabatic eigenfrequencies using empirical models to describe the Lagrangian perturbation of turbulent pressure. 

To this end, \cite{Rosenthal1999} introduced two empirical models. The reduced-$\Gamma_1$ model (RGM), which consists in arguing that the perturbation of turbulent pressure is negligible compared to the perturbation of gas pressure, and the gas-$\Gamma_1$ model (GGM), which consists in considering that perturbations of gas pressure and turbulent pressure are in phase. The authors favored the GGM as it permits them to better reproduce the observed solar frequencies. Most of the above-mentioned works then used the GGM, except for a few \citep[e.g.][]{Jorgensen2019} who considered the RGM guided by the   non-adiabatic calculation by \cite{Houdek2017}. However, the latter was recently challenged by \cite{Schou2020} using the eigenfunctions as inferred directly from 3D numerical simulations. Therefore, the issue of surface effects remains and demands extensive theoretical investigation to gain insight into the physics of the problem.

Hence, the recent progress on surface effects relies on the use of empirical models for which the physics is not understood well enough. In this article, we aim at deriving and then assessing the theoretical validity of the GGM and RGM empirical models. 

\section{Equation governing the turbulent pressure}

When including the mean turbulent pressure in the adiabatic oscillation equations, the Lagrangian perturbation of turbulent pressure is to be prescribed. This is the role of the GGM and RGM models, which are extensively introduced in Appendix~\ref{ggmrgm}. To derive them, a first step is to express the equation for  turbulent pressure. 

\subsection{Averaged equation for turbulent pressure}
\label{aver_pt}

We start with the equation for the averaged Reynolds stresses \citep[e.g.][]{Canuto1997,gatski2013}
\begin{align}
\label{eq_k_1}
\deriv{r_{ij}}{t} &= - r_{ij} \, \partial_k  \widetilde{u}_k - r_{jk} \, \partial_k \widetilde{u}_i  - r_{ik} \, \partial_k \widetilde{u}_j - \partial_k \,\overline{\rho u_i^{\prime\prime} u_j^{\prime\prime} u_k^{\prime\prime}} 
 -\overline{u_i^{\prime \prime} \,\partial_j  P_g}  \nonumber \\
& - \overline{u_j^{\prime \prime} \,\partial_i P_g} 
+ \overline{u_j \, \partial_k \tau_{ik}} + \overline{u_i  \, \partial_k \tau_{jk}} \, , 
 \end{align}
where the overbar denotes the Reynolds average and the tilde denotes the Favre average (see Appendix~\ref{ggmrgm} for a definition), $u_i$ the $i$ component of the velocity field, $P_g$ is the gas pressure (it includes contributions from the radiative field as well as body forces), $\rho$ is the density, $\tau_{jk}$ is the viscous stress tensor, and $r_{ij} \equiv \overline{\rho u_i^{\prime\prime} u_j^{\prime\prime}}$ are the Reynolds stresses where $u_i^{\prime\prime} $ is the $i$-th component of the velocity fluctuation around its Favre average ($\tilde{u}_i$). We also employ the notation $\partial_t \equiv \partial / \partial t$ and $\partial_i \equiv \partial / \partial x_i$, Einstein's notation for repeated indices, and the pseudo-Lagrangian derivative is defined by $\textrm{d}/\textrm{d}t \equiv \partial / \partial t + \widetilde{u}_j \, \partial / \partial x_j$.

We then consider the $rr$ component of \eq{eq_k_1} (where $r$ is the radial coordinate) and identify the Reynolds average with the horizontal average. We obtain
\begin{align}
\label{eq_k_2}
\frac{1}{2} \deriv{P_t}{t} &= - \frac{3 P_t}{2} \, \partial_r  \widetilde{u}_r - \overline{u_r^{\prime \prime}} \,\partial_r \overline{P_g} 
- \partial_r \left( F_{rrr}^{\rm visc} + F_{rrr}^{\rm turb} +  F_r^p \right) \nonumber \\ 
 &+ \overline{P_g^\prime \,\partial_r  u_r^{\prime \prime}}  - \overline{\tau_{rk} \, \partial_k u_r} \, , 
 \end{align}
where $P_t \equiv \overline{\rho u_r^{\prime\prime} u_r^{\prime\prime}}$ is the turbulent pressure, $F_{rrr}^{\rm turb} \equiv \overline{\rho u_r^{\prime\prime} u_r^{\prime\prime} u_r^{\prime\prime}} / 2$, $F_{rrr}^{\rm visc} \equiv - \overline{ u_r \, \tau_{rr}}$, and $F_r^p = \overline{u_r^{\prime \prime} \, P_g^\prime} $. Except for notational differences, \eq{eq_k_2} is strictly equivalent to Eq.~(16b) in \cite{Canuto1997}. 
The first term of the right-hand side of \eq{eq_k_2} is a source of turbulent pressure due to compression produced by radial oscillations. The second term corresponds to the pressure work (or the buoyancy work), which is a source of turbulent pressure in convective regions. The three following terms in brackets correspond to transport terms. Finally, the last two terms are dissipative terms. 

Several assumptions are now needed. The molecular diffusion flux ($F_{rrr}^{\rm visc}$) is neglected compared to the other fluxes appearing in \eq{eq_k_2}. This is justified as we are considering fully developed turbulence with high Reynolds numbers \citep[e.g.][]{Nordlund2009}. The last term of \eq{eq_k_2} is assumed to be proportional to the rate of dissipation of turbulent kinetic energy into heat, \emph{i.e.} $ \overline{\tau_{rk} \, \partial_k u_r} \propto \overline{\tau_{ij} \, \partial_i u_j}$. This standard assumption is made possible because dissipation by molecular forces occurs at almost isotropic small scales \citep[e.g.][]{Pope2000}. 
Finally, the pressure-dilatation term ($\overline{P_g^\prime \,\partial_r  u_r^{\prime \prime}}$) is neglected because it scales as the square of the turbulent Mach number \citep{Sarkar1992} (the maximum is about 0.3 in the Sun). By adopting those approximations, \eq{eq_k_2} reduces to
\begin{align}
\label{eq_k_3}
\deriv{P_t}{t} &= - 3 \, P_t \, \partial_r  \widetilde{u}_r - 2 \, \overline{u_r^{\prime \prime}} \,\partial_r \overline{P_g} 
- \partial_r  F_{r}^{p} - \partial_r  F_{rrr}^{\rm turb} - \frac{2}{3} \, \overline{\rho} \epsilon \, , 
\end{align}
where $\overline{\rho} \epsilon \equiv \overline{\tau_{ij} \, \partial_i u_j}$ is the dissipation rate of turbulent kinetic energy into heat. 

\subsection{Modelling the transport of turbulent pressure}
\label{closure}

The flux of turbulent pressure, $F_{rrr}^{\rm turb}$ in \eq{eq_k_3}, is difficult to model in the uppermost layers of solar-like stars for which the down-gradient approximation fails \citep[see][]{Canuto2009,Kupka2017}. To overcome this issue, we adopt the closure initially proposed by \cite{Canuto1992} \citep[see also][]{Canuto2011} and supported by 3D direct numerical simulations \citep{Kupka2007}. It reads 
\begin{align}
\label{closure_initial}
F_{iir}^{\rm turb}  = c^{-1} \, \frac{k}{\epsilon} \,  F_r^\epsilon \, ,
\end{align}
where $F_{iir}^{\rm turb}$ is the flux of turbulent kinetic energy, $F_r^\epsilon \equiv \overline{\rho \epsilon u_r^{\prime \prime}}$ is the flux of turbulent dissipation, $k \equiv r_{ii}/(2 \overline{\rho})$ is the specific turbulent kinetic energy, and $c$ is a parameter to be specified. 

From \eq{closure_initial}, we then obtain 
\begin{align}
\label{closure_final}
\partial_r F_{rrr}^{\rm turb} = \frac{\alpha}{c} \omega^{-1} \, \partial_r F_r^\epsilon  - \alpha F_{iir}^{\rm turb} \, \partial_r \ln \omega \, , 
\end{align} 
where $\omega \equiv \epsilon / k$ is the turbulence frequency and  
\begin{align}
\alpha \equiv \frac{\partial_r F_{rrr}^{\rm turb}}{\partial_r F_{iir}^{\rm turb}} \, .
\end{align}
The parameter $\alpha$, which can be understood as the degree of anisotropy of the flux of turbulent kinetic energy, will be supposed to be known and obtained directly from the solar 3D numerical simulation (see Sect.~\ref{application}). 

Now, using \eq{closure_final}, \eq{eq_k_3} can be rewritten 
\begin{align}
\label{eq_k_4}
\deriv{P_t}{t} &= - 3 \, P_t \, \partial_r  \widetilde{u}_r - 2 \, \overline{u_r^{\prime \prime}} \,\partial_r \overline{P_g} - \frac{2}{3} \, \overline{\rho} \epsilon - \frac{\alpha}{c} \frac{k}{\epsilon} \, \partial_r F_r^\epsilon \nonumber \\
& - \partial_r F_r^p +  \alpha F_{iir}^{\rm turb} \, \partial_r \ln \omega \, , 
 \end{align}
 where one still needs a prescription for the flux of turbulent dissipation. To do so, we use the equation governing $\epsilon$. A standard procedure consists in constructing a parametric $\epsilon$-equation as is  done in two-equation models (\emph{e.g.} $k-\epsilon$ models). Following this approach one can write \citep[e.g.][]{Pope2000,wilcox2006,gatski2013}
\begin{align}
\label{epsilon_eq}
\overline{\rho} \, \deriv{\epsilon}{t} + \partial_r F_r^\epsilon &=  
- c_1^\epsilon \, \omega \, P_t \, \partial_r  \widetilde{u}_r  - c_2^\epsilon \, \overline{\rho} \epsilon \, \partial_r  \widetilde{u}_r 
- c_3^\epsilon \, \omega \, \overline{u_k^{\prime \prime}} \,\partial_k  \overline{P_g} \nonumber \\
&- c_4^\epsilon \, \overline{\rho} \omega \epsilon \, . 
\end{align}
The first term in the right-hand side of \eq{epsilon_eq} is related to the production of turbulent kinetic energy by compression.  The second term represents for the effect of bulk compressions and expansions onto $\epsilon$ \citep{Coleman1991}. The third term is related to production of turbulent kinetic energy by the buoyancy work, and the last term is related to both the viscous destruction and production due to vortex stretching \citep{gatski2013}.  The coefficients $c_1^\epsilon, c_2^\epsilon,c_3^\epsilon$, and $c_4^\epsilon$ will be discussed in the following section. 

\section{Recovering the gas-$\Gamma_1$ and reduced-$\Gamma_1$ models}

We will now derive an expression for the Lagrangian perturbation of turbulent pressure and make a number of assumptions whose validity will be discussed.  

\subsection{Perturbation of turbulent pressure}
\label{perturb_pt_final}

Using \eq{eq_k_4} together with \eq{epsilon_eq}, we obtain the desired expression for the equation governing turbulent pressure
\begin{align}
\label{eq_k_5}
&\deriv{P_t}{t} = \left(3-\frac{\alpha c_2^\epsilon}{2 c} \Phi - \frac{\alpha c_1^\epsilon}{c} \right) \, \frac{P_t}{\overline{\rho}} \deriv{\overline{\rho}}{t} + \left(\frac{\alpha c_3^\epsilon}{c} - 2\right) \, \overline{u_r^{\prime \prime}} \,\partial_r \overline{P_g} \nonumber \\ 
&+\left(\frac{\alpha c_4^\epsilon}{c}-\frac{2}{3}\right) \, \overline{\rho} \epsilon  
+ \frac{\alpha}{c} \frac{\overline{\rho}}{\omega}  \deriv{\epsilon}{t}  - \partial_r F_r^p +  \alpha F_{iir}^{\rm turb} \, \partial_r \ln \omega \, , 
 \end{align}
where we used the averaged continuity equation (Eq.~\ref{eq_3}), and $\Phi \equiv 2 r_{ii} / P_t$ is the anisotropy factor.

To go further, we assume the following hypothesis:  
\begin{description}
\item [(H1)] the Lagrangian perturbation of the pressure-velocity fluctuations is neglected, \emph{i.e.} $\delta F_r^p = 0$. 
\item [(H2)] the turbulence frequency ($\omega$) is supposed to vary on a length-scale much larger than the length-scale of $F_{iir}$. Accordingly, the second term of the right-hand-side of \eq{closure_final} can be neglected and thus last term of \eq{eq_k_5} vanishes. 
\item [(H3)] the Lagrangian perturbation of $\alpha$ is neglected,  \emph{i.e.} $\delta \alpha = 0$. 
\item [(H4)] the Lagrangian perturbation of the dissipation of turbulent kinetic energy into heat ($\delta \epsilon$) is neglected. 
\item [(H5)] the perturbation of the buoyancy work, $\delta(\overline{u_r^{\prime \prime}} \,\partial_r \overline{P_g})$, is neglected. 
\item [(H6)] the Lagrangian perturbation of density is assumed to be real. 
\end{description}
Perturbing \eq{eq_k_5} and applying H1 to H6, one gets
\begin{align}
\label{eq_deltaPt_1}
&\frac{\delta P_t}{P_t} = \left(3-\frac{\alpha c_2^\epsilon}{2 c} \Phi - \frac{\alpha c_1^\epsilon}{c} \right) \, \frac{\delta \rho}{\rho}  \, , 
 \end{align}
 where $\delta P_t$ and $\delta \rho$ are the pseudo-Lagrangian perturbations of turbulent pressure and density, respectively. Anticipating on the next section, one can already mention that \eq{eq_deltaPt_1} allows one to recover the RGM and GGM models. 
 
 Tracing back the origin of \eq{eq_deltaPt_1} shows that it results from two terms in \eq{eq_k_2}: the source of turbulent pressure due to compression produced by the oscillations and the divergence of the flux of turbulent pressure. 
 It is also important to mention that H1 to H6 are fully consistent with the assumptions made to obtain the equation for the perturbation of gas pressure ($\delta P_g$, see \ref{pgas_classic} and Appendix~\ref{gas_pressure_vs_density}). 
 Moreover, given H1, H3 and H6 and further neglecting the perturbation of the radiative flux, H5 is equivalent to adopting the adiabatic limit (see Appendix~\ref{variation_entropy}). Hence, both equations for $\delta P_t$ (Eq.~\ref{eq_deltaPt_1}) and $\delta P_g$ (Eq.~\ref{pgas_classic}) are compatible with the adiabatic approximation. Conversely, the adiabatic approximation only is not suffisent to derive \eq{eq_deltaPt_1} and \eq{pgas_classic}. 
 
 The first assumption, H1, is essentially equivalent to neglecting the perturbation of the convective flux ($\delta F_r^{\rm conv}$) because $F_r^p$ is proportional to the convective flux as shown by \cite{Canuto1997} (see his Eq.35c). Such an assumption is however not strictly valid because the perturbation of the convective flux does not vanish even in the adiabatic limit \citep[e.g.][]{Sonoi2017}.  H2 is difficult to properly assess because one needs to determine $\epsilon$ and the simplest possible way is to consider $\epsilon \propto k^{3/2} / H_p$ \citep[e.g.][]{Pope2000}, where the dissipation length-scale is assumed to scale as the pressure scale-height ($H_p$). Using, the solar 3D simulation described in \cite{Belkacem2019} one readily finds that H2 is valid near the super-adiabatic peak but not in the quasi-adiabatic regions. 
For H3, it is equivalent to assuming that perturbations of horizontal and vertical contributions of turbulent kinetic energy adjust instantaneously to each other and are thus in phase. A look at the third-order equation on fluxes \citep[see][]{Canuto1997} demonstrates that  the situation is much more complex because many terms are capable to  introduce some redistribution, and in turn phase shifts, when perturbed. Concerning H4 and H5, it consists in neglecting the Lagrangian perturbation of both the turbulent dissipation and the buoyancy work. For mode damping, these two contributions exactly compensate the contribution of turbulent pressure in the limit of a vanishing flux of turbulent kinetic energy and with $\Gamma_3-1=2/3$ (where $\Gamma_3 \equiv \left(\partial \ln T / \partial \ln \rho\right)_s$ ) \citep[e.g.][]{Ledoux58,MAD05}. More recently, \cite{Belkacem2019} demonstrated using the normal modes of a 3D solar hydrodynamical simulation that the perturbation of both terms plays an essential role for the mode damping rates. Hence, as modal surface effects also partly rely on the phase differences between the perturbations of density and turbulent pressure, the impacts of those contributions to surface effects are definitively to be assessed. Finally, H6 assumes that adding the mean turbulent pressure to the hydrostatic equilibrium only introduces a negligible phase shift to the perturbation of density in the adiabatic limit. 

These remarks lead to question the validity of both the RGM and GGM models because, even in the adiabatic limit,  they introduces oversimplifying hypotheses regarding the properties of turbulent convection and mode physics.

\subsection{Application to solar $p$-mode frequencies}
\label{application}

We will now investigate how \eq{eq_deltaPt_1} permits us to recover the GGM and RGM models. A prerequisite is to specify the 
 coefficients $c_1^\epsilon$ and  $c_2^\epsilon$. A commonly accepted value for the first coefficient is  $c_1^\epsilon \simeq 1.44$ \citep[e.g.][]{Pope2000,wilcox2006,gatski2013}. For $c_2^\epsilon$ we adopt the model of \cite{Coleman1991} which gives 
\begin{align}
\label{c2_epsilon}
c_2^\epsilon = \frac{1}{3} \left[1 + 3 n \left(\Gamma_1-1\right) - 2 c_1^\epsilon\right] \, ,
\end{align}
where $n \simeq 0.75$ is the exponent of the viscosity law on temperature. Indeed, \eq{c2_epsilon} aims at accounting for the effect of bulk compressions and expansions onto $\epsilon$ due to the dependence of the viscosity on temperature \citep{Coleman1991}. We note that \eq{c2_epsilon} has been obtained using the rapid distorsion theory \citep[see][for reviews]{Hunt1990,Cambon1993} and under the assumption of adiabatic compression. In the quasi-adiabatic regions of the solar convection zone, those approximations are relatively applicable because the modal period is much shorter than both the typical turn-over time-scale and the thermal time-scale. However, in the super-adiabatic layers, 
 those time scales become of comparable magnitude and the validity of \eq{c2_epsilon} becomes questionable. Therefore, the adopted value of $c_2^\epsilon$ used in this work must be considered a guideline rather than a firm value. 

We thus computed the frequency differences (for radial modes) between the observed frequencies, taken from \citet{Broomhall2009} and \citet{Davies2014}, and theoretical frequencies computed with a classical shooting method. For the latter, we used the RGM, GGM, and the model developed in this work using several values of the closure coefficient $c$ (see Eq.~\ref{closure_initial}). Theoretical frequencies have been obtained by integrating equations \eqs{eq_5}, (\ref{eq_6}), (\ref{pgas_classic}), complemented by either \eq{eq_s_7} for the RGM, \eq{eq_s_8} for the GGM, and \eq{eq_deltaPt_1} for the model proposed in this article. The equilibrium model had been obtained by patching a CESTAM model together with a solar ANTARES 3D simulation \citep[see][for details]{Belkacem2019} using the same methodology as described in \cite{Sonoi2015}. The 3D model is however not exactly solar because the effective temperature is $5750 \pm 18$K and the chemical composition is $(X,Y,Z)=(0.7373,0.2427,0.0200)$ with \cite{GN93} mixture. To allow for a comparison with observed solar frequencies, the 3D model is patched with a CESTAM model with a helium abundance of $0.2485$ and the resulting frequencies are rescaled to match with the standard solar frequency $(GM_\odot/R_\odot^3)^{1/2}/(2\pi) = 99.85537 \, \mu$Hz ($G$ is the gravitational constant, $M_\odot$ the solar mass, and $R_\odot$ the solar radius). Such a procedure is sufficient for our purposes, even if a frequency shift remains at low frequencies, because we are interested in differential effects. From this patched model, all the equilibrium quantities have been inferred. 

\begin{figure}[t]
	\begin{center}
		\includegraphics[width=9.2cm]{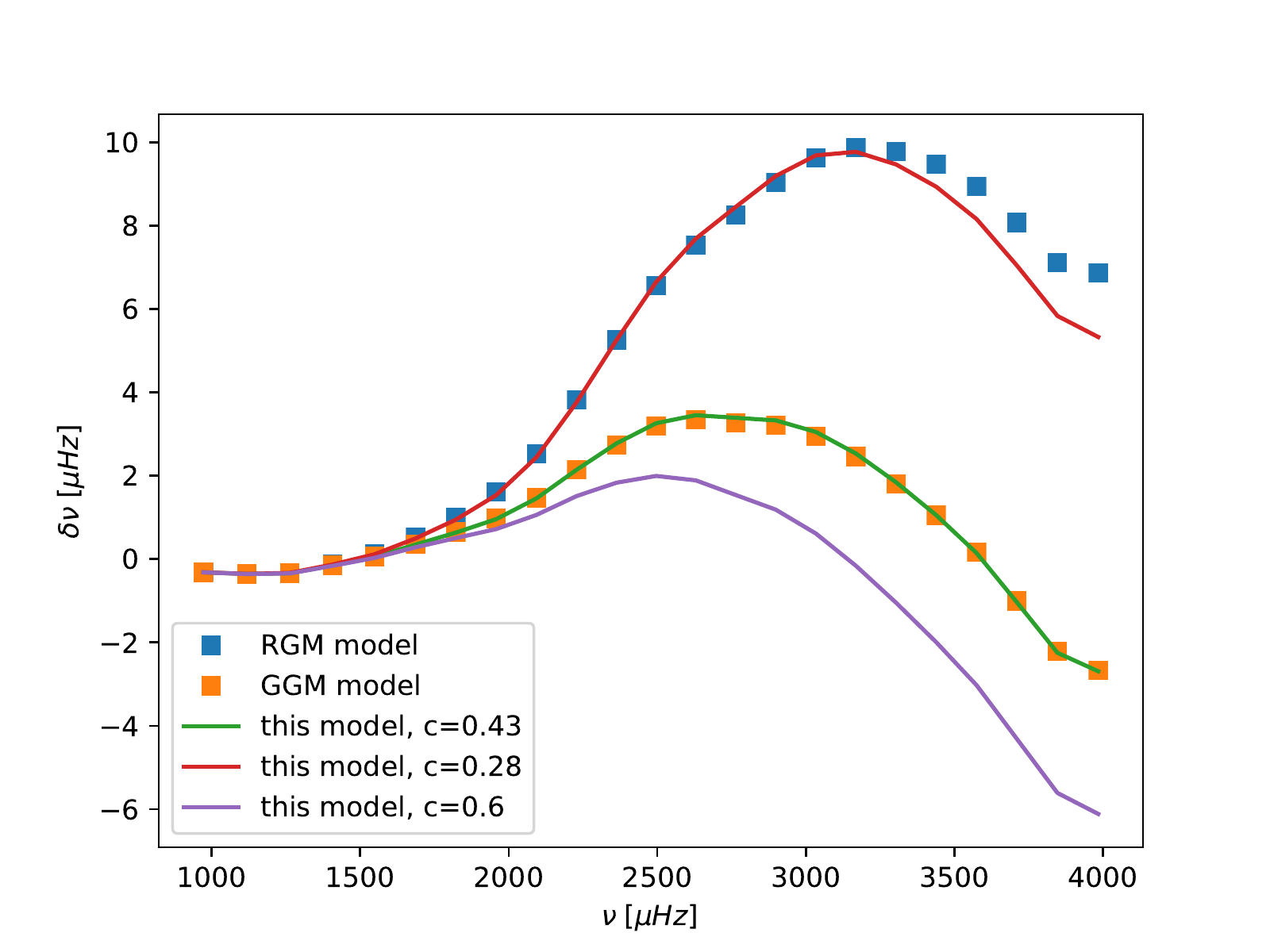}
		\caption{Frequency difference for radial solar modes between the observed frequencies, as inferred by \citet{Broomhall2009} and \citet{Davies2014}, and the modelled frequencies computed using the GGM model, the RGM model, and the model of the present work (see text for details). }
		\label{fig1}
	\end{center}
\end{figure}

Figure~\ref{fig1} shows that both the frequency differences obtained using the GGM and the RGM models can be recovered by adjusting the value of the closure coefficient $c$ introduced in \eq{closure_initial}. 
What guidance for $c$ can we take from existing 3D numerical simulations? A value of $c \approx 0.6$ was derived from four {\em direct numerical simulations}
of fully compressible convection, two of which were discussed in \citet{Kupka2007}. Those simulations were done with the ASCIC3 code \citep{Muthsam1995} 
which is distinguished for such work by not introducing, directly or indirectly, subgrid scale viscosities: dissipation is by explicit, physical viscosity and a very small 
contribution from time integration only. Data were collected for a total of 12 different model configurations (five of which had been discussed in detail in \citealt{Muthsam1995}
and in \citealt{Muthsam1999}). They covered a range in Prandtl number Pr from 0.1 to 1 at a ``zone Rayleigh number'' of $10^5$ to $10^6$. For a horizontal domain 
width eight times the size of this zone depth and with 3 to 4 granules found along each horizontal direction, this implies a ``granulation diameter based Rayleigh number'' 
$\sim 16$ to 50 times larger or  an Ra in the range of $10^6$ to $5 \cdot 10^7$. This yields a product of Ra and Pr in the range of $10^5$ to 
$5 \cdot 10^7$. That (squared) ratio of the thermal diffusion time scale to the buoyancy time scale agrees with results for the upper part of the solar convection zone 
\citep{Kupka2020} despite the convection zones are much more shallow than in the solar case, where additionally ${\rm Pr} \ll 10^{-6}$. 
\citet{Kupka2007} found $c \approx 0.6$ in all cases where sufficient numerical resolution had been ensured, also for cases not shown therein. 
Confirming those results for much lower values of Pr and higher ratio of total flux to radiative flux would be useful. Recalling the discussion in \citet{Belkacem2019} on 
computing the dissipation rate from realistic solar simulations we have to point out here that {\em large eddy simulations}, whether they use hyperviscosity, 
a Smagorinsky-type subgrid scale model, or a Riemann solver, are not the best tools to compute $F_r^\epsilon$ or $\epsilon$: those quantities depend on viscosity
related processes which peak near small multiples of the grid scale. This could heavily bias computations of those quantities by the numerical method used. 
We thus consider conclusions based on low Pr direct numerical simulations for the physical range of ${\rm Ra} \cdot {\rm Pr}$ of interest the safer way to estimate $c$.
In comparison, the ratio $\alpha$ from Eq.~(\ref{closure_final}) can be safely estimated from solar granulation simulations, as the contributions to this quantity peak at length scales resolved in those simulations.

\section{Conclusions}
 
By using the Reynolds and Favre averaged fully compressible Navier-Stokes equations, we have demonstrated that it is possible to develop a model which recovers both the RGM and GGM empirical models. Interestingly enough, this is based on a relation that has been shown to originate from a compensation between the source of turbulent pressure due to compression produced by the oscillations and the divergence of the flux of turbulent pressure. We then showed the RGM and GGM models are compatible with the adiabatic approximation but also imply more drastic and unrealistic physical assumptions regarding turbulence and mode physics. 
  
 Comparison with solar frequencies shows that, while recovering the RGM and GGM, the results are sensitive to the closure coefficients ($c$ in Eq.~\ref{closure_initial} but also  $c_2^\epsilon$ appearing in Eq.~\ref{eq_deltaPt_1}). To consolidate the value of these parameters, only a \emph{direct} 3D numerical simulations could help. Unfortunately, this is still out-of-reach for the Sun due to our current numerical capacities and hence extrapolations from more accessible parameter ranges remain necessary.
 
 It is thus difficult to draw a conclusion on which is the more appropriate to use among the two models. Even worse, given the above-mentioned assumptions which are needed to recover these empirical models, one can safely conclude none of them are firmly physically grounded, even in the adiabatic limit. 
However, one shall still quantify the hypotheses (H1 to H6) on which both RGM and GGM models rely, and more precisely, their individual effect on mode frequencies. This will be made possible by either using a realistic treatment of turbulent convection based on a time-dependent and non-adiabatic treatment or by using normal modes of direct 3D numerical simulations. For the latter, dedicated long-duration simulations (to have a sufficient statistic and to resolve the normal modes) with a large spatial extension (to have a sufficient number of normal modes) need to be computed and this must be done by resolving all spatial scales to obtain an accurate estimate of turbulent dissipation. For the former, a qualitative leap forward is needed because current 1D formalisms based on the mixing-length theory all have their shortcomings among which are free parameters and questionable physical assumptions \citep[see][for details]{Houdek2015}, which prevent them from guaranteeing to properly grasp the physics of the problem. 

\begin{acknowledgements}
      F.~Kupka is grateful to the Austrian Science Fund FWF for support through projects P29172-N and P33140-N and support from European Research Council (ERC) Synergy Grant WHOLESUN \#810218.
\end{acknowledgements}


\appendix

\section{The gas-$\Gamma_1$ (GGM) and reduced-$\Gamma_1$ (RGM) empirical models}
\label{ggmrgm}

When turbulent pressure is included in the mean stratification for computing the classical adiabatic oscillations, it is necessary to determine the perturbation of turbulent pressure and subsequently the perturbation of total pressure \citep[e.g.][]{Rosenthal1999}. To illustrate it let us begin by considering the mass and momentum conservation equations that read 
\begin{align}
\label{eq_1}
&\partial_t \,  \rho + \partial_i \, \left(\rho u_i\right) = 0 \, , \\
\label{eq_2}
&\partial_t \left( \rho u_i \right)+ \partial_k \left(\rho u_i u_k\right) = -\partial_i P_g - \rho g_i + \partial_k \tau_{jk}  \, , 
\end{align}
where $P_g$ is the gas pressure, $g_i$ is the $i$-th component of the gravitational acceleration, $\tau_{jk}$ is the viscous stress tensor, $\rho$ is the density, and $u_i$ the $i$-th component of the velocity field. We also employ the notation $\partial_t \equiv \partial / \partial t$ and $\partial_i \equiv \partial / \partial x_i$ as well as Einstein's notation for repeated indices. 

Equations (\ref{eq_1}) and (\ref{eq_2}) are averaged using both a classical Reynolds average and a density-weighted average \cite[also commonly named to as Favre average, see][]{Favre1969}. For a quantity $X$, the Reynolds average is defined as
\begin{align}
\overline X \equiv X - X^\prime \, , \quad \textrm{with}  \quad \overline{X^\prime} = 0 \, , 
\end{align}
and the Favre average is defined, for a quantity $Y$, by 
\begin{align}
\widetilde{Y} = \frac{\overline{\rho Y}}{\overline Y} \quad \textrm{so that} \quad	Y = \widetilde{Y} + Y^{\prime\prime} \,  \quad \textrm{and}  \quad \overline{\rho Y^{\prime\prime}} = 0 \, .
\end{align}

As  we  consider  a  compressible  flow, using a Reynolds average for density and gas pressure and a Favre average  for the velocity field greatly simplifies the averaged equations \citep[e.g.][]{Canuto1997,Nordlund2001,Belkacem2019}. Applying this procedure for \eqs{eq_1} and (\ref{eq_2}) gives 
\begin{align}
\label{eq_3}
&\deriv{\overline{\rho}}{t}  + \overline{\rho} \, \partial_i \widetilde{u}_i = 0 \, , \\
\label{eq_4}
&\overline{\rho} \deriv{\widetilde{u}_i}{t}
+ \partial_k r _{ik} = -\partial_i \overline{P}_g - \overline{\rho} \,\overline{g_i} \, ,
\end{align}
where $r _{ik} \equiv \overline{\rho \, u_i^{\prime \prime} u_k^{\prime \prime} }$ are the Reynolds stresses (where $u_k^{\prime \prime}$ is the $k$-th component of the velocity fluctuation around its Favre average), the pseudo-Lagrangian derivative is defined by
$\textrm{d}/\textrm{d}t \equiv \partial / \partial t + \widetilde{u}_j \, \partial / \partial x_j$. 
To derive \eqs{eq_3} and (\ref{eq_4}), gravity fluctuations have been neglected and it has been assumed that viscosity does not affect the mean momentum equation and thus the large-scale flow ($\widetilde{u}_j$). It is important to notice that this approximation does not mean that viscous dissipation is neglected, because it appears in the equations governing the turbulent quantities \citep{Canuto1997}. 

We now identify the Reynolds average with the horizontal average so that in \eq{eq_4} only the $rr$ component of the Reynolds stress remains. Turbulent pressure is thus defined by $P_t \equiv \overline{\rho u_r^{\prime \prime} u_r^{\prime \prime}}$. In addition, we split the mean quantities such that $\overline{X} = X_0 + \delta X$ (where $X_0 \equiv <\overline{X}>_t$ is the time-average of $\overline{X}$ and $\delta X$ is the pseudo-Lagrangian perturbation corresponding to the radial oscillations). Therefore, from \eqs{eq_3} and (\ref{eq_4}) we get the desired oscillation equations in the pseudo-Lagrangian frame
\begin{align}
\label{eq_5}
&\frac{\delta{\rho}}{\rho_0}  + \frac{1}{r^2} \, \deriv{}{r} \left( r^2 \xi_r \right) = 0 \, , \\
\label{eq_6}
&\rho_0 \, \sigma^2 \xi_r = \deriv{ \delta P_{\rm tot}}{r} + \delta \rho \, g_0 \, ,
\end{align}
where $\sigma$ is the angular frequency, $\xi_r$ is the radial component of the eigen-displacement ($\widetilde{u}_r = i \omega \, \xi_r$) and $ \delta P_{\rm tot} =  \delta P_{g} +  \delta P_{t}$ is the Lagrangian perturbation of total pressure. In addition, in the adiabatic limit, we also have the thermodynamic relation  
\begin{align}
\label{pgas_classic}
\frac{\delta P_{g}}{P_{g,0}} = \Gamma_1  \frac{\delta \rho}{\rho_0} \, , 
\end{align}
which has been derived by using a number of approximations that are explicitly stated in Appendix~\ref{gas_pressure_vs_density}. 
Therefore, except for the boundary conditions, the system is not closed because we must specify $\delta P_t$ or equivalently $\delta P_{\rm tot}$. Notice that in the following the subscript "$0$" for denoting the temporally and horizontally averaged quantities will be dropped for ease of notation. 

To solve \eqs{eq_5} and (\ref{eq_6}), one needs to specify a prescription for the perturbation of total pressure. It is thus necessary to express the perturbation of turbulent pressure with the perturbation of density so that 
\begin{align} 
\label{def_A}
\frac{\delta P_t}{P_t} = \mathcal{A} \, \frac{\delta \rho}{\rho}  \, , 
\end{align}
where $\mathcal{A}$ is to be determined. Therefore, using \eq{def_A}, one  formally writes
\begin{align}
\label{delta_ptot_generic}
\frac{\delta P_{\rm tot}}{P_{\rm tot}} &= \Gamma_1^{\rm eff} \, \frac{\delta \rho}{\rho} \, , 
\quad 
\textrm{with}
\quad 
\Gamma_1^{\rm eff} \equiv \left[ \Gamma_1 \frac{P_g}{P_{\rm tot}}+\mathcal{A} \frac{P_t}{P_{\rm tot}} \right] \, .
\end{align}

Two empirical models have then been introduced by \cite{Rosenthal1999}. The first is the reduced-$\Gamma_1$ model (RGM), which consists in arguing that $\delta P_t = 0$ or $\mathcal{A}=0$ so that \eq{delta_ptot_generic} becomes
\begin{align}
\label{eq_s_7}
\frac{\delta P_{\rm tot}}{P_{\rm tot}} =  \frac{\delta P_{g}}{P_{\rm tot}} =  \Gamma_1^{\rm eff} \, \frac{\delta \rho}{\rho} \, , 
\quad 
\textrm{where} 
\quad 
\Gamma_1^{\rm eff} = \frac{\Gamma_1 P_g}{P_{\rm tot}} 
\end{align}
is called the reduced $\Gamma_1$. This approximation was initially introduced by \cite{Rosenthal1995} based on the observation that some non-local mixing-length theory shows that density and gas pressure perturbations are almost in phase quadrature with the perturbation of turbulent pressure. The authors therefore consider that the real part of $\delta P_t$ can be neglected. 
The second model is the gas-$\Gamma_1$ model (GGM) and has been introduced by \cite{Rosenthal1999}. It consists in arguing the opposite, i.e. that perturbation of gas pressure and turbulent pressure are in phase. Hence, using $\mathcal{A}=\Gamma_1$ together with \eqs{pgas_classic}, (\ref{def_A}), and (\ref{delta_ptot_generic}), one obtains  
\begin{align}
\label{eq_s_8}
\frac{\delta P_{\rm tot}}{P_{\rm tot}} = \frac{\delta P_t}{P_t} = \frac{\delta P_g}{P_g} = \Gamma_1 \, \frac{\delta \rho}{\rho} \, . 
\end{align}

As can be seen with \eqs{eq_s_7} and (\ref{eq_s_8}), both the RGM and GGM present the major advantage of being easily implemented for computing adiabatic oscillation while including the effect of turbulent pressure, provided that the mean turbulent pressure is prescribed. For GGM it is sufficient to replace the gas pressure by the total pressure in the classical adiabatic oscillation equations while for the RGM one has to also replace $\Gamma_1$ by the reduced $\Gamma_1^{\rm eff}$ given by \eq{eq_s_7}.  

\section{Relation between perturbation of gas pressure and density}
\label{gas_pressure_vs_density}

We follow the derivation proposed by \cite{Rosenthal1999}. They started with the equation governing specific entropy. It reads
\begin{align}
\label{eq_s_1}
\rho T \derivt{s}{t} = - \partial_k F_k^{\rm rad} + \tau_{ik} \partial_k u_i\, ,
\end{align}
where $D/Dt \equiv \partial / \partial t + u_i \partial_i$, $\rho$ is the density, $T$ is the temperature, $s$ is the specific entropy, $F_k^{\rm rad}$ is the $k$-component of the radiative flux, and $\tau_{ik}$ is the viscous stress tensor. 

Equation~(\ref{eq_s_1}) can be recast in terms of gas pressure and density by using the thermodynamic identity
\begin{align}
\label{eq_s_2}
\rho T \derivt{s}{t} = \frac{1}{\left(\Gamma_3-1\right)} \left[ \derivt{P_g}{t} - \frac{\Gamma_1 P_g}{\rho} \derivt{\rho}{t} \right] \, , 
\end{align}
where $P_g$ is the gas pressure, $(\Gamma_3-1) \equiv \left(\partial \ln T /\partial \ln \rho \right)_s$. Then, using \eq{eq_s_2} together with the conservation of mass, \eq{eq_s_1} finally becomes  after averaging 
\begin{align}
\label{eq_s_4}
\deriv{\overline{P}_g}{t} &= - \overline{\Gamma_1 P_g} \, \partial_i \widetilde{u}_i - \overline{\Gamma_1 P_g \, \partial_i u_i^{\prime \prime}}
- \overline{u_i^{\prime \prime} \partial_i P_g} - \overline{\left(\Gamma_3-1\right) \partial_k F_k^{\rm rad}} \nonumber \\
&+ \overline{\left(\Gamma_3-1\right) \tau_{ik} \partial_k u_i} \, , 
\end{align}
which is strictly equivalent to Eq.~(11) of \cite{Rosenthal1999}. Following the same authors, it is assumed that time-varying parts (i.e. the Lagrangian perturbation) of the last four terms of \eq{eq_s_4} are zero. Consequently, the perturbation of \eq{eq_s_4} permits us to recover \eq{pgas_classic}, that is 
\begin{align}
\label{eq_s_5}
\frac{\delta P_g}{P_g} = \Gamma_1\, \frac{\delta \rho}{\rho} \, , 
\end{align}
which is the classical relation for adiabatic oscillations. Note that we assumed $\left<\overline{\Gamma_1 P_g}\right>_t \simeq \left<\overline{\Gamma_1}\right>_t  \left<\overline{P_g}\right>_t$, which is quite an accurate approximation as verified by our solar numerical 3D simulation. 

The assumptions made to derive \eq{eq_s_5} from \eq{eq_s_4}, which consist in assuming that the Lagrangian perturbation of the last four terms of \eq{eq_s_4} are zero, need further discussion. Indeed, 
as recognized by \cite{Rosenthal1999}, those approximations are quite radical. It is nevertheless useful to go a step further and to explain what are the underlying physical assumptions. To that end, let's recast \eq{eq_s_4} in the following form 
\begin{align}
\label{eq_s_6}
\deriv{\overline{P}_g}{t} &= - \Gamma_1 \overline{P_g} \, \partial_i \widetilde{u}_i 
+ \left(\Gamma_1-1\right) \, \overline{u_i^{\prime \prime}} \partial_i \overline{P_g} 
+ \left(\Gamma_1-1\right) \,  \partial_i  \left(\overline{u_i^{\prime \prime} P_g^\prime} \right) \nonumber \\
& - \Gamma_1 \partial_i \left( \overline{P_g \, u_i^{\prime \prime}} \right)
- \left(\Gamma_1-1\right) \, \overline{P_g^\prime \partial_i  u_i^{\prime \prime}}
- \left(\Gamma_3-1\right) \, \partial_k \overline{F_k^{\rm rad}} \nonumber \\
& + \left(\Gamma_3-1\right) \, \overline{ \tau_{ik} \partial_k u_i} \, , 
\end{align}
where, for sake of simplicity and without loss of meaning, we assumed that thermodynamic quantities are time-independent.  To recover \eq{eq_s_5} from \eq{eq_s_6}, the Lagrangian perturbations of the last six terms must be neglected. More precisely; 
\begin{itemize}
\item the perturbation of the buoyancy work ($\overline{u_i^{\prime \prime}} \partial_i \overline{P_g}$), which also appears as a source of turbulent kinetic energy and thus a sink of thermal energy, is considered to be null. 
\item the perturbation of the divergence of the terms $\overline{P_g \, u_i^{\prime \prime}}$ and $\overline{u_i^{\prime \prime} P_g^\prime}$ are set to zero. For the former, considering a perfect gas, it is proportional to the convective (enthalpy) flux because of the relation 
\begin{align}
\overline{P_g u_i^{\prime \prime}} = \mathcal{R} \, \overline{\rho} \, \widetilde{T^{\prime \prime} u_i^{\prime \prime}} \, , 
\end{align}
where $\mathcal{R}$ is the ideal gas constant. Concerning $\overline{u_i^{\prime \prime} P_g^\prime}$, it is also essentially proportional to the convective flux as shown by \cite{Canuto1997} using a polytropic relation. 
Therefore, one can conclude that neglecting the perturbation of those two terms is equivalent to neglect the perturbation of the convective heat flux. 
\item the perturbation of the pressure-strain rate, $\overline{P_g^\prime \partial_i  u_i^{\prime \prime}}$, is neglected. This can be justified as it scales as the squared turbulent Mach number of the rate of dissipation of turbulent kinetic energy into heat \citep{Sarkar1992}. Hence, it can be neglected because we are considering turbulent flow at relatively low turbulent Mach numbers.
\item the perturbations of the radiative flux as well as the dissipation rate of turbulent energy into heat (\emph{i.e.} $\overline{ \tau_{ik} \partial_k u_i}$) are also considered to be negligible. 
\end{itemize}

\section{Averaged equation for specific entropy}
\label{variation_entropy}

To go further, we now aim at inferring the averaged entropy equation. To do so, rather than starting directly with the entropy equation, we consider the equation governing the enthalpy. It reads 
\begin{align}
\label{eq_h_1}
\rho \, \derivt{h}{t} = \derivt{P_g}{t} - \partial_k F_k^{\rm rad} + \tau_{ik} \partial_k u_i\, ,
\end{align}
where $D/Dt \equiv \partial / \partial t + u_i \partial_i$ and $F_k^{\rm rad}$ is the $k-$component of the radiative flux. Now, averaging \eq{eq_h_1} leads to
\begin{align}
\label{eq_h_2}
\overline{\rho} \deriv{\widetilde{h}}{t} &= \deriv{\overline{P}_g}{t}  + \overline{u_k^{\prime \prime}} \,\partial_k  \overline{P_g}
- \partial_k \left( \overline{F}_k^{\rm rad} + F_k^{\rm conv} - \overline{u_k^{\prime \prime} \, P_g^\prime} \right) \nonumber \\
&- \overline{P_g^\prime \, \partial_i u_i^{\prime \prime}}
+ \overline{\tau_{ik} \partial_k u_i} \, ,
\end{align}
where $F_k^{\rm conv} \equiv \overline{\rho h^{\prime \prime} u_k^{\prime \prime}}$ is the $k-$component of the convective flux.  Further using the thermodynamic relation $\overline{\rho} \widetilde{T} \textrm{d}\widetilde{s} = \overline{\rho} \textrm{d}\widetilde{h} - \textrm{d}\overline{P}_g$, valid at leading order, and identifying the Reynolds average to the horizontal average gives 
\begin{align}
\label{eq_h_3}
\overline{\rho} \widetilde{T} \deriv{\widetilde{s}}{t} &= \overline{u_r^{\prime \prime}} \,\partial_r  \overline{P_g}
- \partial_r\left( \overline{F}_r^{\rm rad} + F_r^{\rm conv} - \overline{u_r^{\prime \prime} \, P_g^\prime} \right) 
- \overline{P_g^\prime \, \partial_i u_i^{\prime \prime}} \nonumber \\
&+ \overline{\tau_{ik} \partial_k u_i} \, . 
\end{align}
Neglecting the pressure-strain rate as for the equation governing turbulent pressure and adopting the same notations, one finally obtains 
\begin{align}
\label{eq_h_4}
\overline{\rho} \widetilde{T} \deriv{\widetilde{s}}{t} &= \overline{u_r^{\prime \prime}} \,\partial_r  \overline{P_g}
- \partial_r\left( \overline{F}_r^{\rm rad} + F_r^{\rm conv} - F_r^p \right) + \overline{\rho} \epsilon \, . 
\end{align}

Perturbing \eq{eq_h_4} thus permits us to derive an expression for the perturbation of the buoyancy work. It reads 
\begin{align}
\label{eq_h_5}
\delta \left(\overline{u_r^{\prime \prime}} \,\partial_r  \overline{P_g}\right) =  i \sigma \overline{\rho} \widetilde{T} \delta s 
+ \delta \left[ \partial_r\left( \overline{F}_r^{\rm rad} + F_r^{\rm conv} - F_r^p \right)\right] - \delta \left(\overline{\rho} \epsilon\right) \, . 
\end{align}
Using the same approximations as described in Sect.~\ref{perturb_pt_final} to derive \eq{eq_deltaPt_1} immediately leads to $\delta s = 0$. 

\end{document}